\documentclass[a4paper,11pt]{article}
\usepackage{pos}
\usepackage{multicol}

\title{EUSO-SPB2 Cosmic Ray Searches and Observations}
 \ShortTitle{EUSO-SPB2 Cosmic Rays}
 \author*[a]{G. Filippatos}
 \affiliation[a]{University of Chicago, USA}
 \emailAdd{gfil@uchicago.edu}
 \onbehalf{for the JEM-EUSO collaboration} 

\abstract{
	The Extreme Universe Space Observatory on a Super Pressure Balloon 2 (EUSO-SPB2) flew in May of 2023, marking an important step towards the observation of ultra-high-energy cosmic rays (UHECR) and neutrino-induced showers from space. The ultimate goal of this endeavor is to complement ground-based detectors and achieve unprecedented exposure and nearly uniform full-sky coverage at the highest energies, thereby enabling charged particle astronomy and enriching the multi-messenger approach to high-energy astrophysics and astroparticle physics. As a pathfinder to the POEMMA mission (Probe Of Extreme Multi-Messenger Astrophysics), EUSO-SPB2 flew two distinct cameras at the focus of two Schmidt telescopes, one made of multi-anode photomultiplier tubes (MAPMTs), looking towards the nadir for fluorescence light detection, the other made of Silicon photomultipliers (SiPMs), looking towards the limb of the Earth for direct Cherenkov light detection. The flight was terminated prematurely due to a failure in the balloon, and thus no showers were detected in the fluorescence mode. However, several lower-energy (PeV scale) cosmic-ray events were observed in the Cherenkov channel. The data collected by both telescopes also confirmed the pertinence and maturity of the technology. We will report on the mission's cosmic ray results, and lessons learned for future balloon and satellite missions, notably the POEMMA Balloon with Radio (PBR), currently under development.
}

\ConferenceLogo{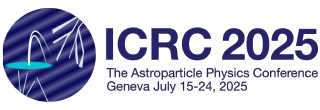}

\FullConference{39th International Cosmic Ray Conference (ICRC2025)\\
 15–24 July 2025\\
Geneva, Switzerland\\}
\begin{document}
\maketitle

\section{Introduction}

Understanding the origin and nature of ultra-high-energy cosmic rays (UHECRs) and neutrinos remains one of the most compelling challenges in astroparticle physics. These particles, with energies exceeding $10^{18}$ eV, carry unique information about the most energetic processes in the universe, yet their extremely low flux requires novel observational strategies to achieve statistically significant measurements. Ground-based detectors such as the Pierre Auger Observatory \cite{auger} and the Telescope Array \cite{TelescopeArray} have made considerable progress but are fundamentally limited in aperture and sky coverage. To overcome these limitations, space-based observatories have been proposed to complement terrestrial efforts by vastly increasing exposure and enabling full-sky coverage with a single instrument and nearly uniform exposure \cite{COLEMAN2023102819}. 

The Extreme Universe Space Observatory on a Super Pressure Balloon 2 (EUSO-SPB2) represents a critical step toward this vision. Flown in May 2023, EUSO-SPB2 served as a technology pathfinder and scientific demonstrator for future space missions, most notably the Probe of Extreme Multi-Messenger Astrophysics (POEMMA) \cite{POEMMA}. The mission deployed two complementary instruments aboard a high-altitude balloon: a fluorescence telescope employing multi-anode photomultiplier tubes (MAPMTs) to observe nitrogen fluorescence from EeV scale extensive air showers (EAS) \cite{ADAMS2025103046}, and a Cherenkov telescope using silicon photomultipliers (SiPMs) to detect direct Cherenkov light from cosmic ray or $\nu_{\tau}$ induced PeV scale air showers near the limb of the Earth \cite{Trinity}.  

The flight was completed after less than 37 hours afloat due to a technical failure of the balloon. Despite not reaching the expected minimum flight duration, data were recorded, which validated the performance of the instruments in the near-Earth observing environment. These critical data\footnote{downloaded via next generation telemetry at $100\times$ the originally planned bandwidth} are invaluable for the development of future missions.

\begin{figure}[h!]
  \centering
  \includegraphics[width=0.47\textwidth]{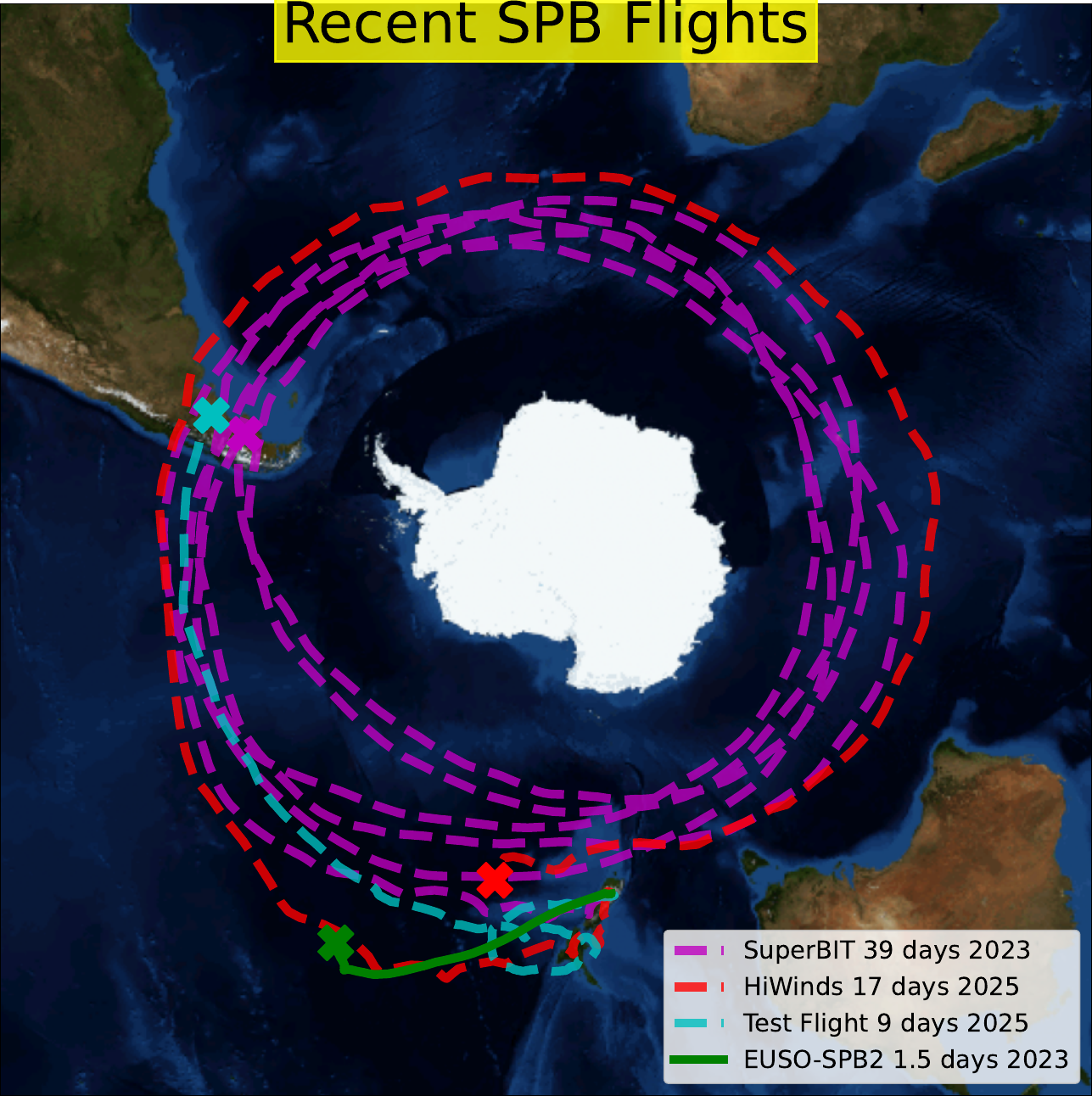}
  \caption{Recent NASA Super Pressure Balloon flights from W\={a}naka NZ. The 18.8~million~ft$^3$ (532,000~m$^3$) balloon is designed to suspend a 5,500~lbs (2,500~kg) payload at 110,000~ft (33.5~km) altitude.}
\end{figure}
\newpage
\section{Fluorescence Measurements}

The fluorescence telescope (FT) was built with the heritage of two previous balloon missions \cite{EUSO-Balloon,SPB1}. The focal surface was made up of 108 MAPMTs, for a total of 6,912 channels. Each channel was capable of counting single photons, with a double pulse resolution of $\sim$10~ns and an integration time of 1050~ns. The currents from groups of 8 adjacent channels were integrated and digitized with the same cadence, providing sensitivity to very bright signals that could saturate the photon counting channel \cite{Miyamoto:2023vb}. The MAPMTs are grouped in fours to form elementary cells (ECs). The MAPMTs along with accompanying ASICs and HVPS are potted in a gelatinous compound to prevent electrostatic discharge due to the high voltage ($\sim$1000V) operating in the near-vacuum of the stratosphere. Groups of nine ECs were connected to an FPGA board to form a photodetection module (PDM). A novel trigger was developed and worked on a per-PDM basis \cite{SPB2Trigger}. When one PDM triggered, all three were read out. The three PDMs sat at the focus of a 1~m diameter entrance aperture Schmidt telescope with a $36^\circ\times12^\circ$ field of view (FoV). The end-to-end efficiency of the detector was $\sim$15\%. The FT was fixed rigidly to the gondola and pointed nadir, which minimized the energy threshold for detecting UHECRs.

The pre-flight expectation was that roughly one UHECR event would be detected in the fluorescence channel for every 10 hours of moonless observation with a peak energy sensitivity around 3~EeV \cite{Filippatos:2021b5}. These expectations were validated by a field campaign at the Telescope Array Black Rock Mesa FD site during August of 2022 \cite{Kungel:2023C6}.
\begin{figure}[h!]
 \centering
 \includegraphics[width=\textwidth]{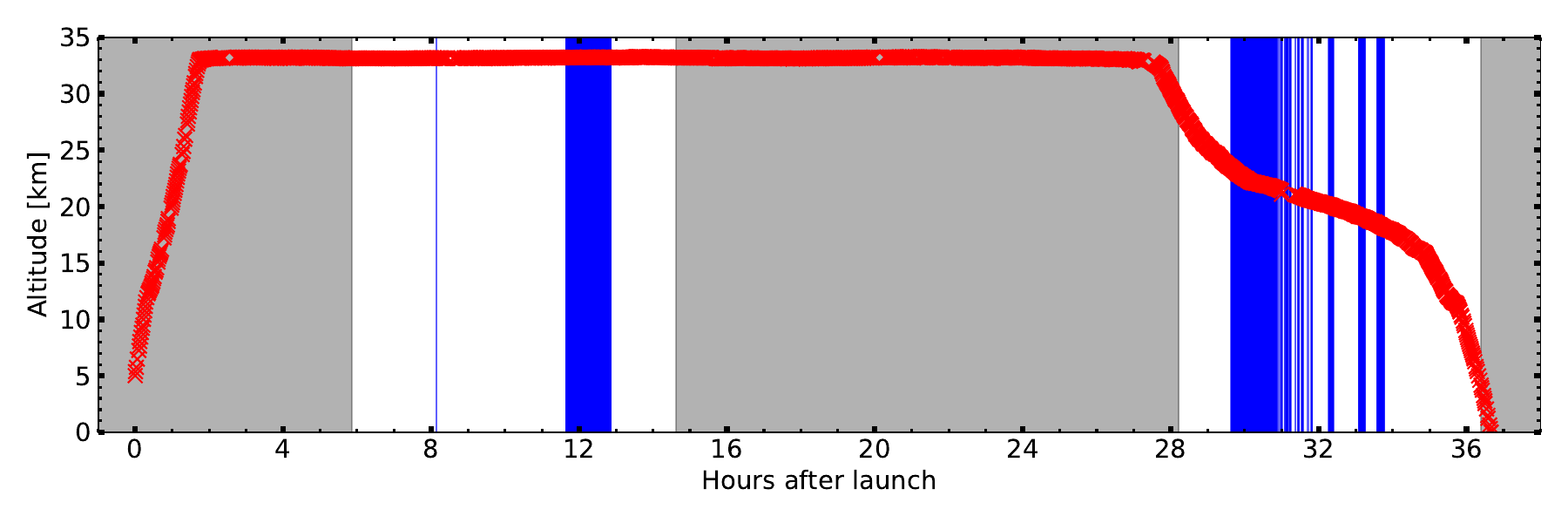}
 \caption{Flight profile with payload altitude shown in red. The gray regions represent periods where the background light was too high to safely observe due to either the sun or the moon, measured by an independent low-voltage photodiode. The blue regions represents periods where the DAQ system was operating normally and the data was downloaded to the ground.}
 \label{figs:ft_led}
\end{figure}

In order to verify the data acquisition system (DAQ) in-situ, a set of two monitoring LEDs were flashed every 16 seconds. The LED system and the PDMs were completely independent, allowing the system to serve as a check of the trigger and to verify that the DAQ was not over-saturated. In order to quantify the "active" period of observation, the periods where consecutive LED flashes are recorded can be used. In addition to verifying the DAQ system was operating normally, these periods represent data that was downloaded to ground. The times the system was "active" are shown in Figure \ref{figs:ft_led}. The total observing time was just under 3 hours. The majority of the observing time occurred on the second night while the payload was descending. The first night was primarily dedicated to commissioning, with the active periods representing times when $\sim$30-50\% of the focal surface, and therefore FoV, were operating normally. On the second night, the full camera was operational. The periods without data are the result of data that was recorded, but not downloaded before the end of the flight.

\begin{figure}[h!]
  \centering
  \includegraphics[width=\textwidth]{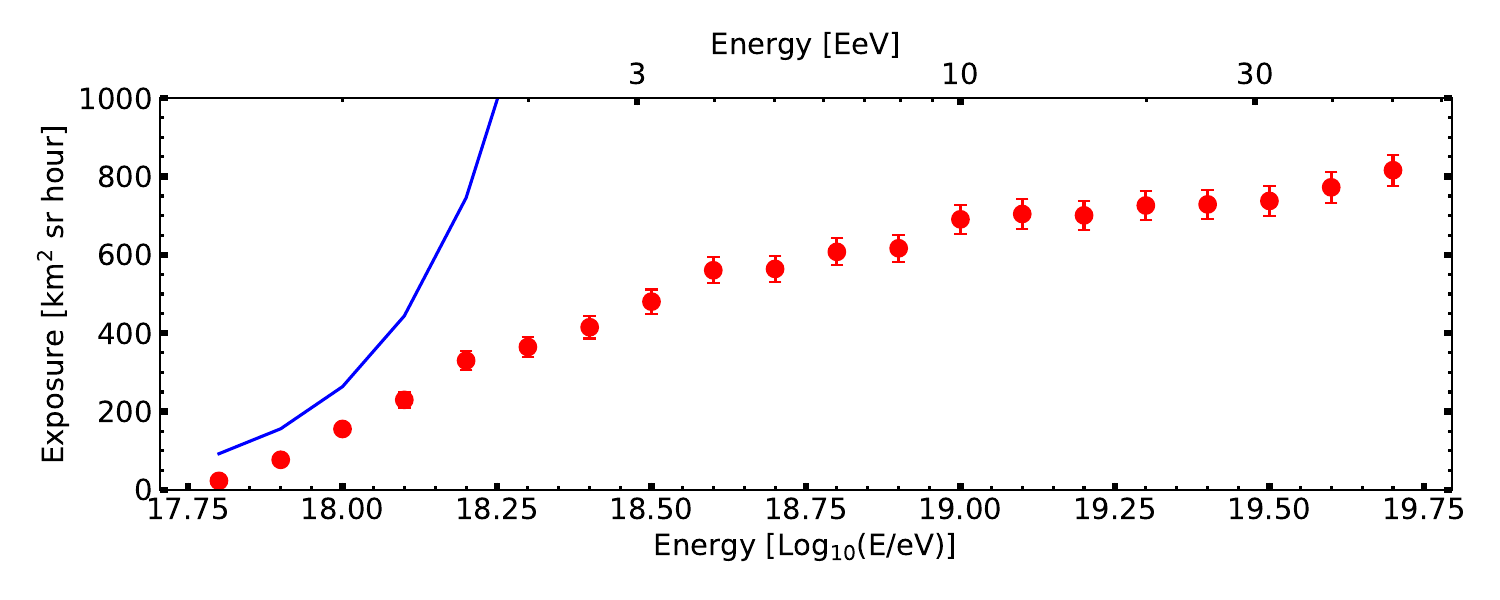}
  \caption{Exposure achieved by the FT over the course of the two night flight, based on extensive end-to-end simulations (red points) and the exposure required to "expect" greater than 1 event above a given energy based on the energy spectrum reported by Auger \cite{PhysRevD.102.062005} (blue line). Methodology described in detail in \cite{ADAMS2025103046}.}
  \label{figs:exposure}
 \end{figure}

Due to the nature of the observation, the aperture of the FT is not trivially defined. This is because there are many EAS geometries that cross the FoV without landing beneath the detector. One way the aperture can be estimated is by isotropically throwing a large number of simulated EAS with core locations covering a large area (R=100~km) below the detector. By comparing the number of EAS that trigger to the total number thrown, the aperture can be estimated. The response of the detector is highly energy dependent, due to higher energy EAS triggering in a larger number of geometries. The aperture also depends on the altitude of the payload. The higher the payload the larger the volume of atmosphere observed, and the higher the energy threshold for triggering due to being further away. In order to account for these effects, 24 million EAS were simulated using the EUSO-OffLine framework \cite{euso-offline}. The achieved exposure of the instrument based on these simulations is shown in Figure \ref{figs:exposure}.

 Accounting for the altitude of observation, the expected number of events based on the energy spectrum reported by Auger \cite{PhysRevD.102.062005} observed over the flight is 1.25. This estimate does not account for clouds, due to the cloud monitoring infrared (IR) camera not operating normally during the descent. No EAS event was found in the $\sim$$10^5$ downloaded events, consistent with this expectation.

 While the downloaded data did not include any UHECR candidates, several important features were observed. These include the presence of clouds aligning with the IR camera (on the first night of the flight) and a linear increase of the background with increasing lunar elevation. Further, the operation of the instrument in the near-space environment provided insight into the detector stability. In particular, the types of events that triggered the readout are different from those which do on the ground, possibly due to the radiation exposure at the altitude of the balloon.
 
\section{Cherenkov Measurements}
The Cherenkov Telescope (CT) was a first-of-its-kind instrument. The camera in the focus of this telescope consisted of 32 SiPM matrices for a total of 512 channels. The analog signal was pre-amplified before being digitized by a modified ASIC developed for the General Electronics for Time projection chambers (AGET) system \cite{POLLACCO201881}, which digitized the signal in 10~ns time bins with 512 bins per readout. The electronics were actively cooled by a custom liquid cooling solution. Similarly to the FT, the CT was a 1~m diameter Schmidt telescope with a smaller $12^\circ\times6^\circ$ FoV. Two noticeable differences between the FT and CT were: 1.$)$ the use of bifocal optics in the CT, with the incident light split into two adjacent spots on the focal surface in order to reject signals from charged particles interacting directly in the SiPMs and 2.) the capability of changing the elevation of the telescope from $+3.5^\circ$ to $-13.5^\circ$ relative to horizontal. The fully assembled CT was tested on the ground, in Utah in 2022 and in New Zealand in 2023 immediately prior to launch, where it observed bifocal signals consistent with the Cherenkov signal expected from EAS.

Similar to the FT, the first night of the flight was primarily dedicated to commissioning the CT. On the second night, the CT observed primarily below-the-limb searching for signals from $\nu_\tau$ induced air showers \cite{SPB2Neutrinos25}. For $\sim$45 minutes of the flight, the CT was pointed above-the-limb where it was sensitive to the Cherenkov light from cosmic ray induced EAS \cite{PhysRevD.104.063029}. After taking time to adjust trigger thresholds accordingly, the detector was sensitive to Cherenkov signals for roughly 37 minutes.

\begin{figure}[h!]
  \centering
  \includegraphics[width=0.7\textwidth]{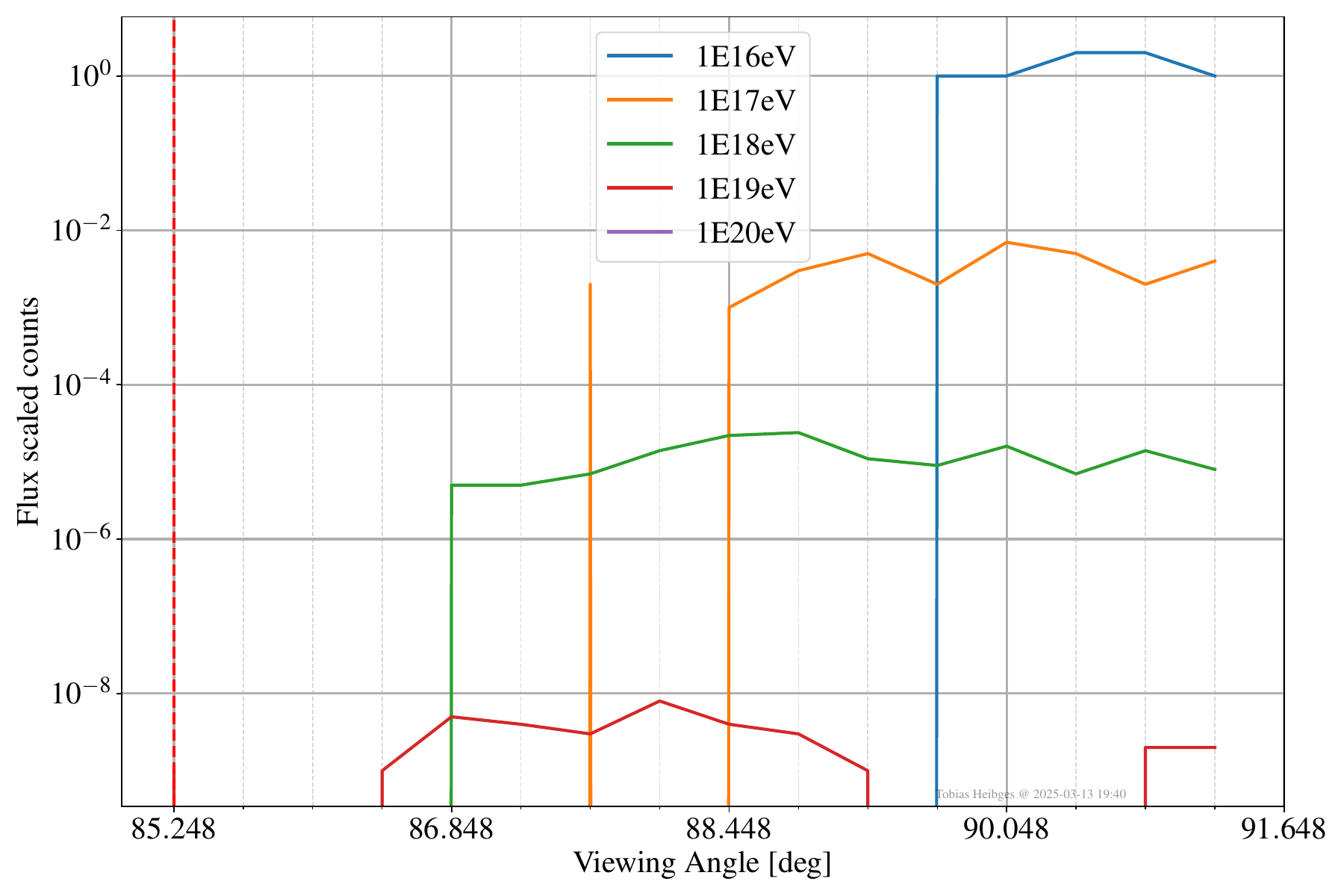}
  \caption{Expected viewing angle distribution, across the vertical FoV of the CT based on simulated showers. The vertical dashed red line represents the direction of the limb, the vertical grey grid lines represent the vertical FoV of a pixel, and the right edge of the plot corresponds to the
top of the CT’s FoV. Taken from \cite{TobiasThesis}}
  \label{figs:ct_angles}
 \end{figure}

 The energy threshold and aperture can be estimated based on simulations preformed using EASCherSim and the EUSO-OffLine framework. These simulations suggest that the turn-on energy for the CT is around $10^{16}$~eV. Accounting for the achieved exposure of the flight, one would expect that all observed events are below 100~PeV. The simulated acceptance for different simulated energies and viewing angles is shown in Figure \ref{figs:ct_angles}. 

 \begin{figure}[h!]
  \centering
  \includegraphics[width=\textwidth]{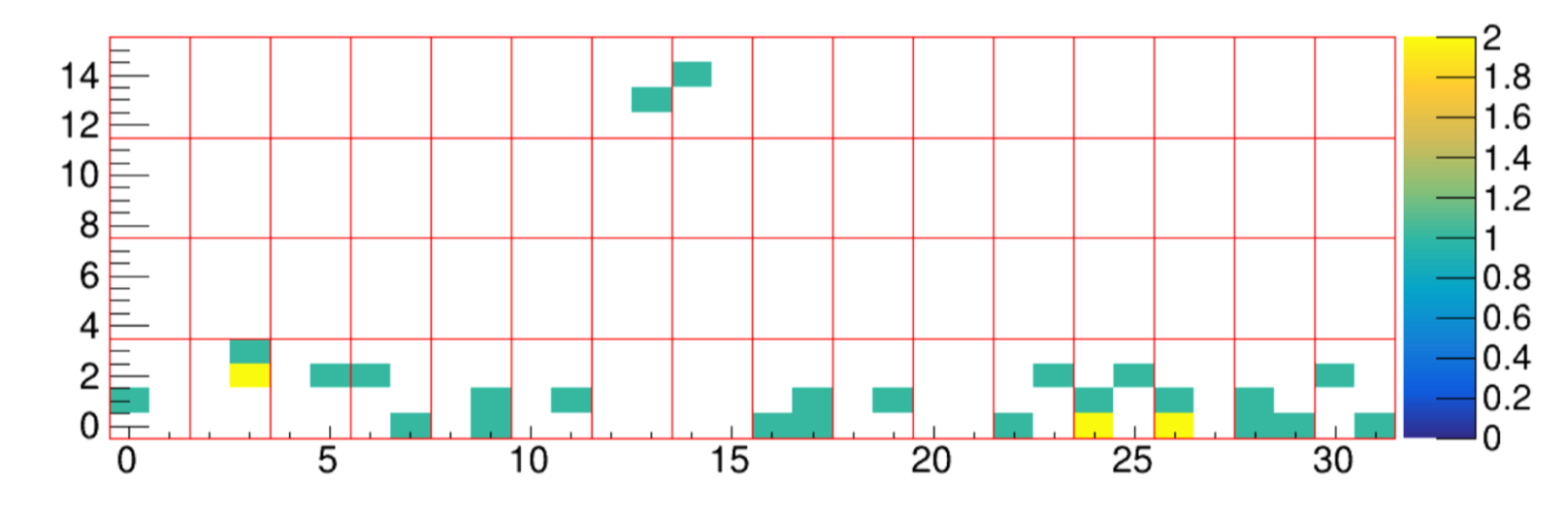}
  \caption{Trigger map of high amplitude bifocal events (both triggered pixels shown per event). Taken from \cite{ElizaThesis}}
  \label{figs:ct_pix}
 \end{figure}

 In total, 14 high-amplitude bifocal events were observed during the period the telescope was pointed above-the-limb, after applying cuts including removing "hot pixels" etc. High-amplitude events here are defined as events with greater than 70 photoelectrons (PEs) in the triggering pixel. This is consistent with the expected flux, and aperture based on simulations. Further validating the simulations is the location of the candidate events within the FoV. Due to the observing geometry, the atmosphere acts as an energy filter. Events that come from closer to the limb of the Earth traverse more atmospheric depth, and the optical light produced experiences more attenuation as a result. Therefore these events require more Cherenkov light to be produced in order to be detected. The locations of all 14 high-amplitude bifocal events is shown in Figure \ref{figs:ct_pix}, where it can be seen that 13 of the events were recorded in the bottom row of SiPM matrices, consistent with this effect. 

 \begin{figure}[h!]
  \centering
  \includegraphics[width=0.9\textwidth]{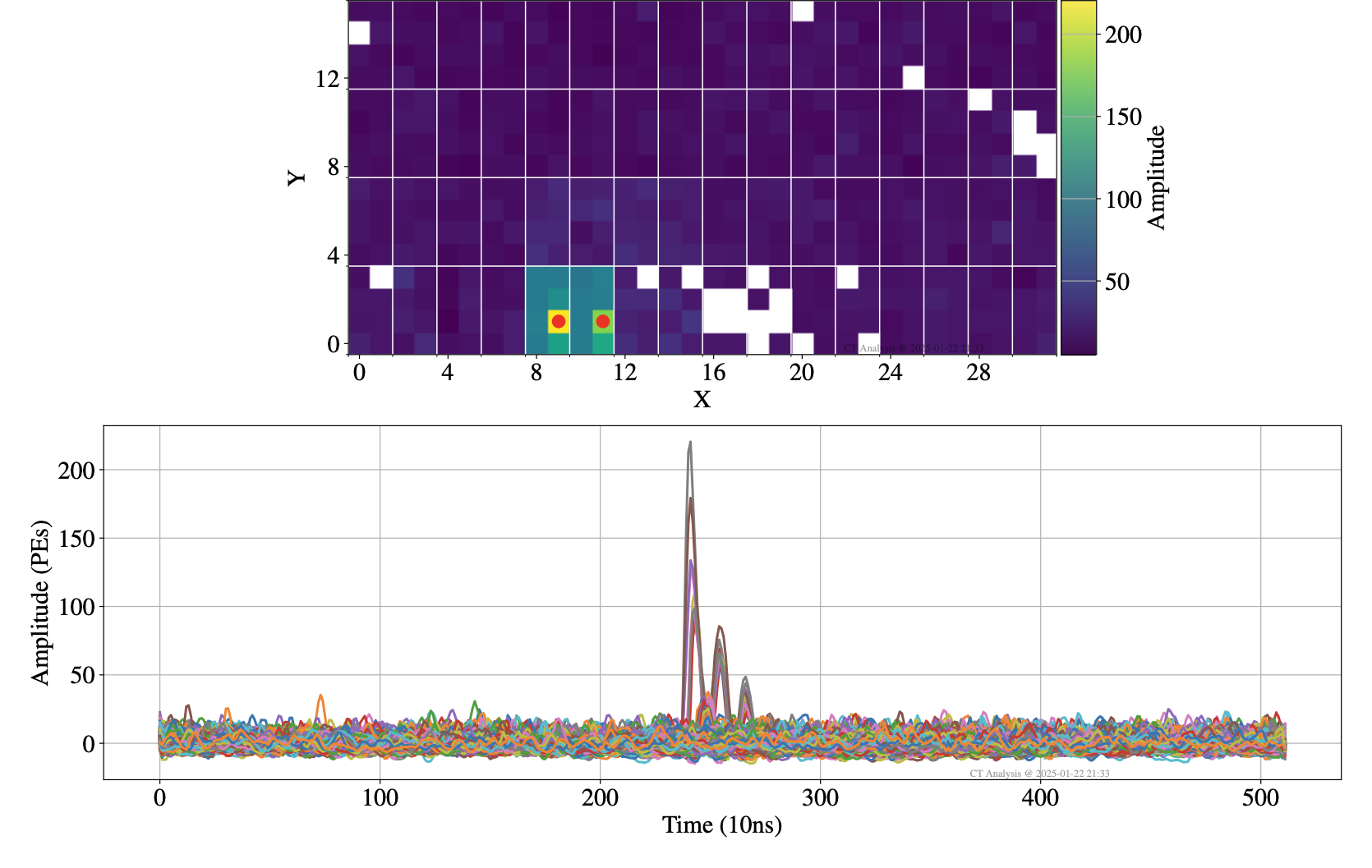}
  \caption{\cite{TobiasThesis}}
  \label{figs:ct_example}
 \end{figure}

 An example of one of these high-amplitude bifocal events is shown in Figure \ref{figs:ct_example}. The signal is split in two pixels separated by the designed 12~mm. Comparing to simulations, this event is consistent with a 10 PeV air shower. A precise estimation of its energy is not possible, as the Chernkov signal is proportional to both the energy of the air shower and the position of the detector relative to the shower axis. The upcoming POEMMA Balloon with Radio mission aims to break this degeneracy by measuring showers in low frequency radio. 

\section{Outlook}
The EUSO-SPB2 mission represents a pivotal step in the development of space-based observation techniques for UHECRs and neutrinos. Despite a significantly shortened flight duration of less than two days, the mission achieved several critical milestones. The FT, while not detecting any UHECRs, successfully validated the end-to-end functionality of its hardware and trigger systems in a near-space environment, and provided key insights into background behavior and detector stability. Importantly, the flight confirmed that the instrument’s performance aligns well with pre-flight expectations and simulations, which may serve as a benchmark for future missions.

The CT succeeded in detecting several bifocal events consistent with the expected signal from PeV-scale cosmic ray air showers. The spatial distribution and characteristics of these events matched predictions from simulations. These measurements provide the first flight-proven demonstration of this novel detection technique. These findings bolster confidence in the feasibility of detecting such events from near-space altitudes and inform refinements for future designs.

Looking ahead, the lessons learned from EUSO-SPB2 will directly inform the POEMMA Balloon with Radio (PBR) mission, which aims to expand on this platform by addressing known limitations, increasing observation duration, and incorporating complementary radio detection. The results from EUSO-SPB2 underscore the maturity of the instrumentation and the viability of balloon-based pathfinders in advancing the goals of full-sky, space-based astroparticle observatories.

\section*{Acknowledgements}
This work was partially supported by Basic Science Interdisciplinary Research Projects of RIKEN and JSPS KAKENHI Grant (22340063, 23340081, and 24244042), by the Italian Ministry of Foreign Affairs and International Cooperation, by the Italian Space Agency through the ASI INFN agreement EUSO-SPB2 n. 2021-8-HH.0 and its amendments and through the ASI-INAF agreement n.2017-14-H.O, by NASA award 11-APRA-21730058, 16-APROBES16-0023, 17-APRA17-0066, NNX17AJ82G, NNX13AH54G, 80NSSC18K0246, 80NSSC18K0473, 80NSSC19K0626, and 80NSSC18K0464 in the USA, by the French space agency CNES, by Slovak Academy of Sciences MVTS JEM–EUSO, by National Science Centre in Poland grant 2020/37/B/ST9/01821, as well as VEGA grant agency project 2/0132/17, and by State Space Corporation ROSCOSMOS and the Interdisciplinary Scientific and Educational School of Moscow University ``Fundamental and Applied Space Research"
This research used resources of the US National Energy Research Scientific Computing Center (NERSC), the DOE Science User Facility operated under Contract No. DE-AC02-05CH11231. 
We acknowledge the contributions of the JEM-EUSO collaboration to the EUSO-SPB2 project.
We thank the Telescope Array collaboration for graciously allowing us to use their facilities for field tests.
\bibliographystyle{JHEP-nt}
\bibliography{refs}

{\Large\bf Full Authors list: The JEM-EUSO Collaboration}

\begin{sloppypar}
{\small \noindent
M.~Abdullahi$^{ep,er}$              
M.~Abrate$^{ek,el}$,                
J.H.~Adams Jr.$^{ld}$,              
D.~Allard$^{cb}$,                   
P.~Alldredge$^{ld}$,                
R.~Aloisio$^{ep,er}$,               
R.~Ammendola$^{ei}$,                
A.~Anastasio$^{ef}$,                
L.~Anchordoqui$^{le}$,              
V.~Andreoli$^{ek,el}$
% ...
}
\end{sloppypar}
\vspace*{.3cm}

{ \footnotesize
\noindent
%
% ...
$^{cb}$ Universit\'e de Paris, AstroParticule et Cosmologie (CNRS), Paris, France\\
% ...
$^{ef}$ Istituto Nazionale di Fisica Nucleare (INFN), Sezione di Napoli, Naples, Italy\\
% ...
$^{ei}$ Istituto Nazionale di Fisica Nucleare (INFN), Sezione di Roma Tor Vergata, Rome, Italy\\
% ...
$^{ek}$ Istituto Nazionale di Fisica Nucleare (INFN), Sezione di Torino, Turin, Italy\\
$^{el}$ Universit\`a di Torino, Dipartimento di Fisica, Turin, Italy\\
% ...
$^{ep}$ Gran Sasso Science Institute (GSSI), L’Aquila, Italy\\
$^{er}$ Istituto Nazionale di Fisica Nucleare (INFN), Laboratori Nazionali del Gran Sasso, Assergi, Italy\\
% ...
$^{ld}$ University of Alabama in Huntsville, Huntsville, AL, USA\\
$^{le}$ City University of New York (CUNY), Lehman College, Bronx, NY, USA\\
% ...
}
\end{document}